\documentclass[sigconf,natbib=True,anonymous=false]{acmart}

\usepackage{booktabs} 
\usepackage{algorithm}
\usepackage{algorithmic} 

\usepackage{subfigure}
\usepackage{graphics}
\usepackage{ifthen}
\usepackage{latexsym}
\usepackage{CJK}
\usepackage{amsfonts}
\usepackage{ifthen}
\usepackage{caption}
\usepackage{graphicx}
\usepackage{float}
\usepackage{multirow}
\usepackage{subfigure}
\usepackage{balance}
\usepackage{subfigure}
\usepackage{graphics}
\usepackage{enumerate}
\usepackage{verbatim}
\usepackage{graphicx}
\usepackage{float}
\usepackage{booktabs}
\usepackage[hang,flushmargin]{footmisc}
\usepackage{caption}
\usepackage{bbold}
\usepackage{color}
\usepackage{xspace}

\usepackage{listings}
\usepackage{xcolor}
\usepackage{multicol}
\usepackage{adjustbox}
\usepackage{calc}
\usepackage{enumitem}
\usepackage{amsmath}

\usepackage{booktabs}    
\usepackage{makecell}    
\usepackage{multirow}    
\usepackage{graphicx}    

\newcommand{\paratitle}[1]{\vspace{1.5ex}\noindent\textbf{#1}}

\newcommand{\ignore}[1]{}

\newcommand{\baby}{\textsc{VQL}\xspace}
\AtBeginDocument{%
  }

\begin{document}

\title{VQL: An End-to-End Context-Aware Vector Quantization Attention for Ultra-Long User Behavior Modeling}

\author{Kaiyuan Li}
\email{likaiyuan03@kuaishou.com}
\affiliation{%
  \institution{Kuaishou Technology}
  \city{Beijing}
  \country{China}
}

\author{Yongxiang Tang}
\email{tangyongxiang@kuaishou.com}
\affiliation{%
  \institution{Kuaishou Technology}
  \city{Beijing}
  \country{China}
}
\author{Yanhua Cheng}
\email{chengyanhua@kuaishou.com}
\affiliation{%
  \institution{Kuaishou Technology}
  \city{Beijing}
  \country{China}
}
\author{Yong Bai}
\email{baiyong@kuaishou.com}
\affiliation{%
  \institution{Kuaishou Technology}
  \city{Beijing}
  \country{China}
}
\author{Yanxiang Zeng}
\email{zengyanxiang@kuaishou.com}
\affiliation{%
  \institution{Kuaishou Technology}
  \city{Beijing}
  \country{China}
}
\author{Chao Wang}
\email{wangchao32@kuaishou.com}
\affiliation{%
  \institution{Kuaishou Technology}
  \city{Beijing}
  \country{China}
}
\author{Xialong Liu}
\email{zhaolei16@kuaishou.com}
\affiliation{%
  \institution{Kuaishou Technology}
  \city{Beijing}
  \country{China}
}
\author{Peng Jiang}
\email{jiangpeng@kuaishou.com}
\affiliation{%
  \institution{Kuaishou Technology}
  \city{Beijing}
  \country{China}
}

\renewcommand{\shortauthors}{Kaiyuan Li et al.}

\begin{abstract}
In large-scale recommender systems, ultra-long user behavior sequences encode rich signals of evolving interests. While extending sequence length generally improves predictive accuracy, directly modeling such sequences in production is often infeasible due to strict latency and memory constraints. Existing solutions fall into two categories: (1) \textbf{top-$k$ retrieval}, which discards $L-k$ interactions and can lose the majority of attention mass when the sequence length $L \gg k$; and (2) \textbf{encoder-based compression}, which preserves sequence coverage but tends to over-compress and fails to incorporate critical context such as temporal gaps and target-aware signals. Consequently, neither approach achieves an optimal balance between low-loss compression, context awareness, and online efficiency.

In this work, we propose \textbf{VQL}, an end-to-end, context-aware \textbf{V}ector \textbf{Q}uantization Attention framework for ultra-\textbf{L}ong user behavior modeling, with three key innovations: 
(1) \textbf{Key-only quantization} — quantizing only attention keys while keeping values intact; we prove that, by softmax normalization, the resulting attention-weight error is independent of sequence length $L$, and that the codebook loss directly supervises the key quantization error. This design also enables caching of quantized keys, yielding $L$-free inference cost. 
(2) \textbf{Multi-scale quantization} — partitioning the $H$ attention heads into $G$ groups, each trained with its own quantizer, and applying quantization within groups to further suppress error while keeping online complexity nearly constant. 
(3) \textbf{Efficient context injection} — most contextual features are natively supported by VQL, with static contexts (e.g., item category, modality) directly integrated and relative position handled via a separable temporal kernel. All injections preserve codebook-based caches, keeping results $Q$-free and inference latency unchanged.

We evaluate VQL on three large-scale datasets (KuaiRand-1K, KuaiRec, TMALL). Experimental results show that VQL consistently outperforms strong baselines, achieving superior accuracy while substantially reducing inference latency. These results establish VQL as a new state-of-the-art in balancing accuracy and efficiency for ultra-long sequence recommendation.

\end{abstract}



\maketitle

\section{Introduction}

\begin{figure}
    \centering
    \includegraphics[scale=0.5]{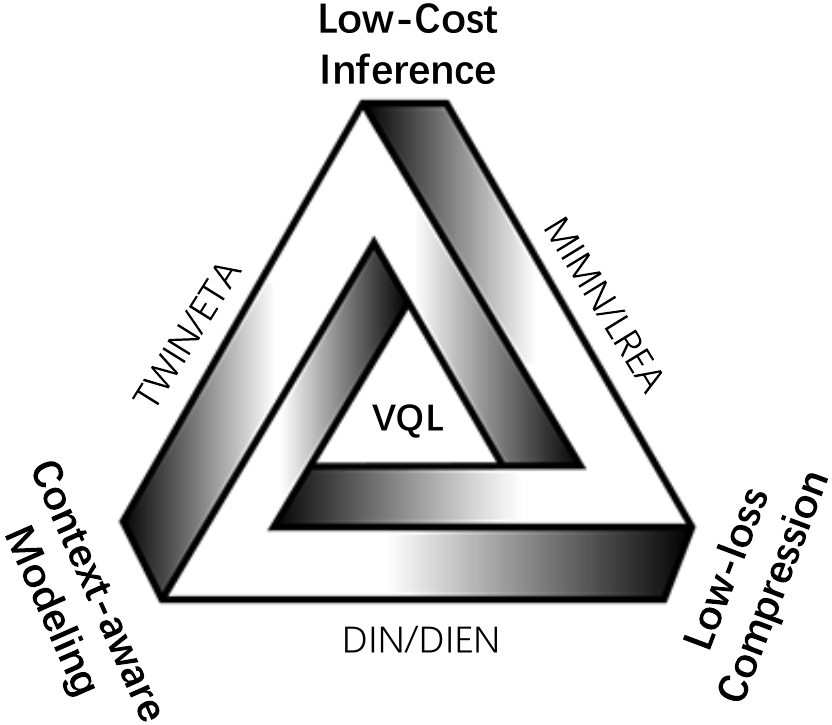}
    \caption{This figure illustrates the balance between low-cost inference, low-loss compression, and context-aware modeling in long-sequence modeling methods. Our proposed \baby can more effectively balance all three aspects.}
    \label{fig:tradeoff}
\end{figure}
Sequential user modeling is a core component of modern recommender systems, supporting tasks such as click-through rate (CTR) and conversion rate (CVR) prediction in e-commerce, as well as engagement prediction in short-video platforms. A common paradigm is target-aware attention, exemplified by models such as DIEN\citep{dien} and DIN\cite{DBLP:conf/kdd/ZhouZSFZMYJLG18}. Empirical evidence consistently shows that extending the length of user behavior sequences significantly enhances the modeling of long-term user interests~\cite{twin_v2,CIKM'20/SIM}. However, directly processing ultra-long sequences in production is infeasible due to strict constraints on online latency and memory.

Existing solutions fall into two categories:
(1) \textbf{Top-$k$ retrieval}: selects a small subset of behaviors via target-aware matching~\cite{CIKM'20/SIM,KDD'23/TWIN}, but discards $L-k$ items, losing substantial attention mass when $L \gg k$.
(2) \textbf{Encoder-based compression}: condenses the entire sequence into fixed-length representations~\cite{lrea,KDD'19/MIMN}, preserving coverage but often over-compressing and ignoring fine-grained context such as temporal gaps and target-awareness.

Neither class effectively balances low-loss compression, context-awareness, and online efficiency - a trade-off summarized in Fig.~\ref{fig:tradeoff}. While models such as DIN and DIEN naturally incorporate context information and preserve the original attention structure, their computational cost scales linearly with sequence length $L$, making them impractical for online inference.
Top-$k$ retrieval methods such as TWIN\cite{KDD'23/TWIN} and ETA\cite{arxiv'21/ETA} can compress sequences down to $k$, but as discussed earlier, they suffer from significant information loss when handling ultra-long sequences.
Compression-based approaches such as MIMN\cite{KDD'19/MIMN} and LREA\cite{lrea} avoid explicit training–inference mismatch errors, yet struggle to capture context signals such as target-awareness. This motivates our work.

In this paper, we introduce \textbf{VQL}, an end-to-end context-aware Vector Quantization Attention framework for ultra-long user behavior modeling. VQL addresses the above limitations with four key advantages:
\begin{enumerate}[leftmargin=4mm]
    \item \textbf{K-only quantization:}  
    VQL compresses only the keys while keeping values intact, avoiding value-side distortion and preserving the fidelity of the original attention structure.  

    \item \textbf{L-Free caching:}  
    By exploiting the one-hot extraction property, all key–value aggregates can be precomputed offline into $L$-independent caches, eliminating online dependence on sequence length $L$.  

    \item \textbf{E-Tight error bound:}  
    The output error is provably bounded in direct proportion to the quantization error, and crucially independent of $L$. To further reduce this error, we adopt Grouped Vector Quantization (GVQ), which leverages multiple small codebooks to enhance representational capacity without inflating cache size.  Codebook and commitment losses directly supervise quantization quality, ensuring tighter bounds.

    \item \textbf{Q-Free context injection:}  
    VQL natively supports the injection of diverse context features (e.g., static attributes, temporal signals) in a query-free manner.  
    Context kernels, such as separable temporal kernels, are integrated without enlarging the codebook, ensuring that cached representations remain query-independent.  
\end{enumerate}

Extensive experiments on three large-scale datasets (KuaiRand-1K, KuaiRec, TMALL) show that VQL consistently outperforms strong baselines.  
In addition, we design multiple caching strategies to flexibly trade off memory footprint and latency under different deployment settings, further improving practicality at industrial scale.

\begin{table*}[t]
\centering
\renewcommand{\arraystretch}{1.2}
\footnotesize
\begin{tabular}{@{}l*{8}{c}@{}}
\toprule
\multirow{2}{*}{Model} & \multicolumn{6}{c}{Compression Attributes} & \multicolumn{2}{c}{Computational Complexity} \\
\cmidrule(lr){2-7} \cmidrule(lr){8-9}
 & \makecell{Method} & \makecell{Content} & Time-aware & End-to-End & \makecell{Supersived} & \makecell{Cache-able} & \makecell{Training} & \makecell{Inference} \\
\midrule
SIM       & top-k search    & KV & No  & No  & No & No  & $O(B\log(M) + Bkd)$    & $O(B\log(M) + Bkd)$ \\
ETA       & top-k search    & KV & Yes & Yes & No & No  & $O(BLm\log(d) + BLm + Bkd)$ & $O(BLm + Bkd)$ \\
TWIN      & top-k search    & KV & Yes & Yes & No & Yes  & $O(BLd + Bkd)$ & $O(BLd + Bkd)$\\
TWIN-V2   & semantic cluster; top-k search & KV & Yes & No  & No & Yes & $O(BTd +Bkd)$   & $O(BTd +Bkd ) $ \\
LREA      & linear weight  & KV & No  & Yes & No & Yes &$O(BLkd+Bkd)$  & $O(Bkd)$ \\
\midrule
VQL & semantic cluster & \textbf{K}
& \textbf{Yes} & \textbf{Yes} & \textbf{Yes} & \textbf{Yes} 
& $O(BLkd + Bkd)$ & \textbf{$O(Bkd)$} \\
\bottomrule
\end{tabular}
\caption{This figure compares compression methods and computational complexity. Here, \( B \) represents the number of candidate items for each request, \( m \) denotes the number of hashes employed by ETA, \( L \) indicates the lengths of the original user behavior sequences, and \( k \) refers to the lengths of the retrieved and compressed user behavior sequences, respectively. \( M \) is the size of the attribute inverted index in the SIM, \(T \)  is the number of cluster centers for TWIN-V2, and \( d \) is the model's hidden size.} 
\label{tab:comp}
\end{table*}

\section{Related Work}
\label{sec:related-work}

\paratitle{User Modeling}. 
Ranking models are a fundamental component of recommendation systems, designed to accurately predict the relevance of candidate items to users. These models primarily focus on tasks such as Click-Through Rate (CTR)~\citep{DBLP:conf/cikm/ShengZZDDLYLZDZ21, DBLP:conf/kdd/WuBCRXHDZ22, DBLP:conf/cikm/ZhangCXBHDZ22, he2014practical, gai2017learning, chang2023pepnet} and Conversion Rate (CVR)~\citep{DBLP:conf/kdd/ChanZHBSLHLJXZ23, DBLP:conf/aaai/YangLHZZZT21, DBLP:conf/kdd/GuSFZZ21} prediction. 

Early ranking models primarily relied on hand-crafted features combined with traditional machine learning algorithms, such as Logistic Regression (LR) and Factorization Machines (FM)~\citep{DBLP:conf/icdm/Rendle10, 10.1145/2959100.2959134}. These methods required significant manual feature engineering to capture user-item interactions, which was often labor-intensive and limited in its ability to model complex relationships.

With the advent of deep learning, researchers have explored the paradigm shift from manual feature engineering to automatic feature interaction using neural networks~\citep{DBLP:conf/recsys/Cheng0HSCAACCIA16, DBLP:conf/ijcai/GuoTYLH17, DBLP:conf/kdd/LianZZCXS18, DBLP:conf/ijcai/XiaoY0ZWC17, DBLP:conf/www/WangSCJLHC21, DBLP:conf/kdd/ZhouZSFZMYJLG18, DBLP:conf/aaai/LyuDHR20, DBLP:conf/kdd/XiaEPBWGJFZZ23, DBLP:journals/corr/abs-2411-09852, zhang2024scenario, wang2017deep}. These neural network-based models have demonstrated their ability to automatically learn complex user-item relationships, enhancing the accuracy and effectiveness of recommendation systems.

\paratitle{Long Sequence Modeling}. Recent advancements in user interest modeling for ranking tasks have significantly benefited from the integration of user behavior data, yielding impressive results. Early approaches focused on utilizing deep neural networks to automatically identify user interests from short-term activities, employing models such as recurrent neural networks (RNNs)~\citep{DBLP:journals/corr/HidasiKBT15, DBLP:conf/aaai/ZhouMFPBZZG19}, target-attention mechanisms~\citep{DBLP:conf/kdd/ZhouZSFZMYJLG18, DBLP:conf/aaai/LyuDHR20, zhou2019deep}, and transformers~\citep{DBLP:conf/kdd/XiaEPBWGJFZZ23, DBLP:journals/corr/abs-2411-09852, sun2019bert4rec, kang2018self}. While these methods effectively capture short-term user behaviors, understanding long-term behavior is essential for tracking the evolution of user interests over time. Memory-based techniques~\citep{DBLP:conf/sigir/RenQF0ZBZXYZG19, KDD'19/MIMN} utilize memory networks to encapsulate long-term interests.

Subsequent approaches, such as SIM~\citep{CIKM'20/SIM} and UBR4CTR~\citep{Sigir'20/UBR4CTR}, have introduced a two-stage cascaded framework. This framework first retrieves the most relevant behaviors from long-term data and then performs complex modeling based on these retrieved behaviors. Building on the SIM framework, numerous studies have further refined the GSU to enhance performance~\citep{CIKM'22/SDIM, arxiv'21/ETA, KDD'23/TWIN, CIKM'24/TWINv2, arxiv'24/MARM, arxiv'24/DARE}. 

Recently, some efforts have been made to abandon this modeling paradigm with impressive results. For example, \cite{lrea} achieved remarkable outcomes by compressing user behavior sequences end-to-end using linear weights. To better understand the differences between existing long-term modeling work and our proposed VQL, Table \ref{tab:comp} shows the differences among various long-term sequence modeling approaches.

Existing retrieval-based and encoder-compression approaches degrade notably as the sequence length grows. In particular, retrieval discards a large fraction of user interactions, while encoder compression tends to over-smooth representations. We empirically validate this in our scaling experiments (Sec. \ref{sec:scaling}), where these baselines suffer a clear drop in both AUC and LogLoss. By contrast, our Key-only VQ achieves consistent gains as the sequence extends, confirming its robustness to ultra-long sequences.

\section{Preliminaries}
In this section, we primarily introduce the fundamental concepts of our task.
\subsection{Task Definition}
Click-through rate (CTR) prediction is a core task in online advertising and recommender systems. Its goal is to estimate the likelihood that a user clicks on a specific item. The CTR prediction task involves learning a function \( f(u, i) \) that represents the predicted probability of user \( u \) clicking on item \( i \). The CTR prediction process can be formalized as:
\begin{equation}
    f(u, i) = \sigma\left(\phi(u, i)\right),
\end{equation}
where \( \phi(u, i) \) denotes the combined feature representation of user \( u \) and item \( i \), and \( \sigma(x) \) is the sigmoid function.

The model is trained to minimize the binary cross-entropy loss:
\begin{equation}
    \mathcal{L}_{\text{Rec}} = -\frac{1}{N} \sum_{j=1}^{N} \left[ y_j \log\left(f(u_j, i_j)\right) + (1 - y_j) \log\left(1 - f(u_j, i_j)\right) \right],
\end{equation}\label{eq:rec_loss}
where \( N \) is the number of training samples, and \( y_j \in \{0, 1\} \) is the binary label indicating whether the user clicked the item.
\subsection{Vector Quantization (VQ)}
Vector Quantization (VQ) is a classic technique for compressing high-dimensional vectors into a finite set of representative prototypes, known as a codebook.  
Formally, given a learnable codebook $C = \{c_j\}_{j=1}^N \in \mathbb{R}^{N \times d}$, each input vector $x \in \mathbb{R}^d$ is mapped to its nearest codeword $\hat{x} = c_{z}$, where $z = \arg\min_j \|x - c_j\|_2$.  
The mapping can be equivalently expressed using a one-hot assignment matrix $\Delta \in \{0,1\}^{1 \times N}$ such that
\begin{equation}
    \hat{x} = \Delta C .
\end{equation}

To enable end-to-end training, VQ is optimized with the standard codebook loss and commitment loss:
\begin{equation}
\mathcal{L}_{\text{VQ}} 
= \|\mathrm{sg}[x] - \hat{x}\|_2^2 
+ \beta \, \|x - \mathrm{sg}[\hat{x}]\|_2^2 ,
\end{equation}\label{eq:vq_loss}
where $\mathrm{sg}[\cdot]$ denotes stop-gradient and $\beta$ controls the strength of the commitment term.  
The first term pulls codewords toward encoder outputs, while the second term encourages encoder outputs to commit to their assigned codewords, preventing codebook collapse.

\section{Methodology}
\label{sec:method}

\begin{figure*}
    \centering
    \includegraphics[width=1\linewidth]{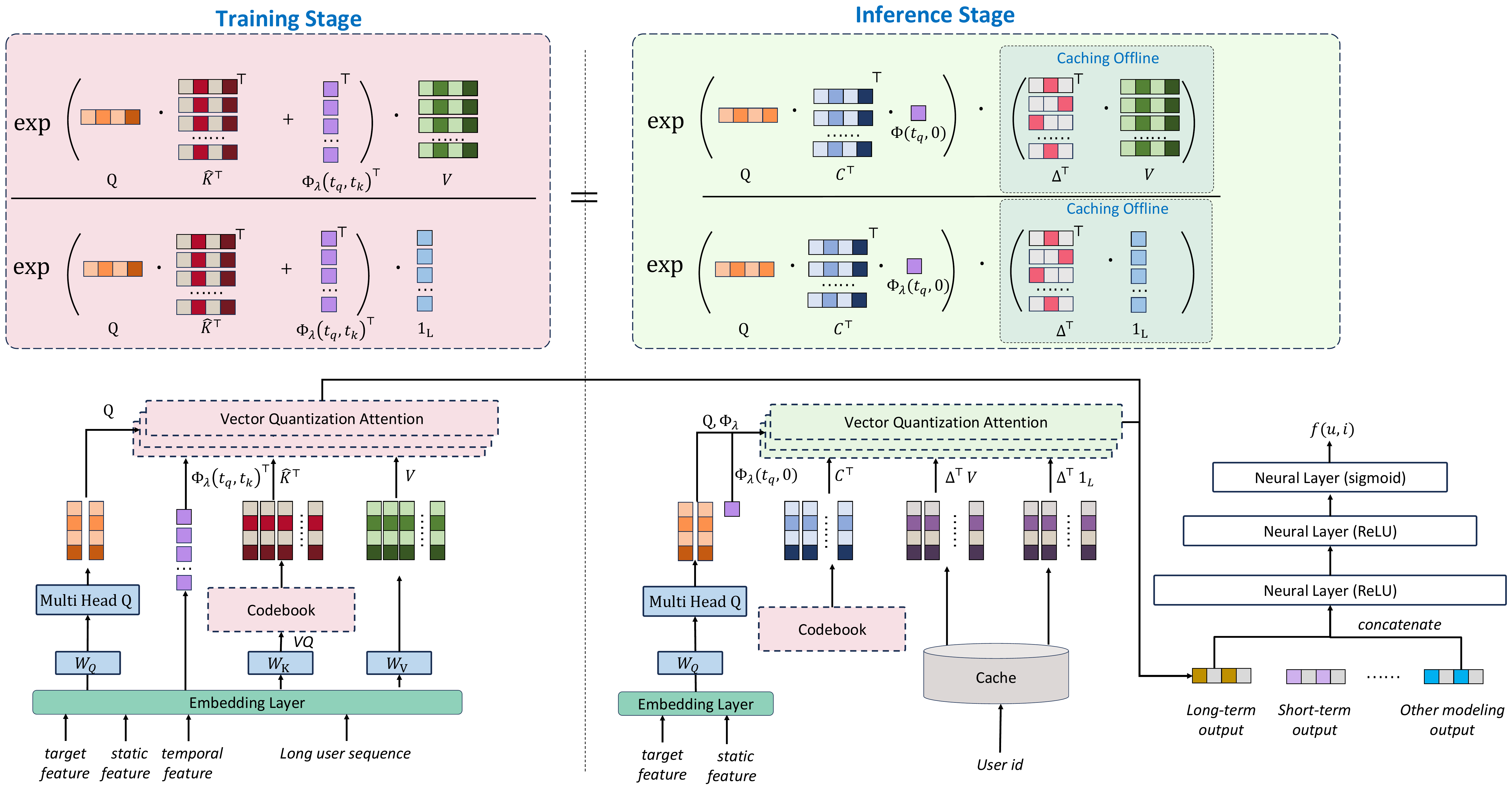}
    \caption{Overview of the proposed \textbf{VQL} framework. 
    The left panel shows the training process with key-only quantization and cache construction, while the right panel shows inference using the same cached results. 
    Training and inference are mathematically equivalent, differing only in implementation. 
    For illustration, we present the case with a single group ($G{=}1$) and a single temporal kernel ($M{=}1$).
    }
    \label{fig:placeholder}
\end{figure*}

We present \textbf{VQL} for ultra-long user behavior modeling in industrial recommenders. 
Our design follows three principles: 
(i) key-only vector quantization that preserves value fidelity while being directly supervised by a codebook loss; 
(ii) an $L$-free, cacheable attention form derived from a one-hot extraction identity so online cost does not grow with sequence length; and 
(iii) \emph{Q-free} context injection. Static attributes and separable scalar signals (e.g., timegaps) are injected \emph{after} one-hot extraction so that cached representations remain query-independent, without affecting the normal caching mechanism.

An overview of the framework is illustrated in Fig.~\ref{fig:placeholder} and we develop these ideas in order.  
First, we revisit scaled dot-product attention and establish notation.  
Next, we introduce key-only vector quantization and prove the one-hot extraction identity that yields the cacheable attention form, together with an $L$-independent output error bound that is tightened by codebook/commitment losses.  
We then extend to Grouped Vector Quantization (GVQ), which uses multiple small codebooks on column-wise groups—akin to decoupling queries and KV groups as in grouped-query attention—to increase representational capacity.
Finally, we describe a cache-compatible mechanism for injecting static and dynamic context (e.g., categorical attributes, separable scalar kernels for temporal or scene signals) after one-hot extraction so that cached representations remain query-independent.

\subsection{Dot-Product Attention Revisited}  
Given a query $q\in\mathbb{R}^{d}$, we define $Q = q / \sqrt{d}$ to absorb the scaling factor for easy reading.  
Let keys $K\in\mathbb{R}^{L\times d}$ and values $V\in\mathbb{R}^{L\times d}$, the scaled dot-product attention becomes
\begin{equation}
\label{eq:att}
o(Q) \;=\;
\frac{\exp\!\big(QK^\top\big) \; V}
     {\exp\!\big(QK^\top\big) \; \mathbf{1}_L},
\end{equation}
where $\exp(\cdot)$ acts element-wise and $\mathbf{1}_L$ is the $L$-dimensional all-ones vector.  
Directly forming $QK^\top$ is $O(Ld)$ per query; caching $K,V$ per user is $O(L(d{+}d))$ memory.  
Our goal is to transform \eqref{eq:att} into a form where the \emph{online} cost is independent of $L$.

\subsection{Key-only Vector Quantization Attention}
\label{subsec:vq-onehot}
For the standard attention structure, previous top-$k$ methods often truncate the sequence, thereby discarding $L{-}k$ keys and values.  
Compressing both $K$ and $V$ risks accuracy losses that compound in the output when $L \gg k$.  
Inspired by \cite{vq_transformer,vq_vae}, we \emph{quantize only the keys}, preserving the raw values to avoid value-side distortion while still reducing the attention computation.  
We find that key-only quantization brings substantial benefits, such as enabling $L$-free offline caching that, unlike traditional methods, does not increase online inference cost, as well as providing an $L$-free quantization error bound that can be further tightened with a supervised loss.  
We will elaborate on and prove these findings in the following subsections.

\subsubsection{Key-only Vector Quantization.}
Let $C=\{c_j\}_{j=1}^N\in\mathbb{R}^{N\times d}$ be a learnable codebook ($N\!\ll\!L$).  
Each key $k_i$ is replaced by its nearest codeword $\hat{k}_i=c_{z_i}$, where $z_i\!\in\!\{1,\dots,N\}$.  
We collect the assignments into a one-hot matrix $\Delta\in\{0,1\}^{L\times N}$ (one $1$ per row), so that:
\begin{equation}
\label{eq:KQ}
\hat{K} \;=\; \Delta C .
\end{equation}
The codebook and assignments are optimized end-to-end with the standard codebook and commitment losses (Eq.~\ref{eq:vq_loss}), which directly reduce $\|K-\hat{K}\|_F$.  
Next, we will prove the $L$-free caching property and the $L$-free error bound advantage.

\subsubsection{L-free Cacheable attention form.}
The core algebraic tool that enables caching is the following identity.

\begin{lemma}[One-hot extraction]
\label{lem:onehot}
For any element-wise function $\phi$, any matrix $U\in\mathbb{R}^{m\times N}$, and any $W\in\{0,1\}^{N\times L}$ whose columns are one-hot (exactly one 1 per column), we have
\begin{equation}
\label{eq:onehot-phi}
\phi(U)\, W \;=\; \phi(UW).
\end{equation}
\end{lemma}

\noindent\textit{Proof.}
Let the $j$-th column of $W$ have its 1 at row $s_j$. Then for any row $r\in\{1,\dots,m\}$,
\[
[\phi(U)W]_{rj}
= \sum_{n=1}^N \phi(U_{rn}) W_{nj}
= \phi(U_{r s_j})
= \phi([U W]_{rj}),
\]
proving \eqref{eq:onehot-phi} entry-wise for all $r,j$. \hfill $\square$

\begin{align}
\exp\!\big(Q\hat{K}^\top  \big)
&= \exp\!\big(QC^\top\Delta^\top\big)
\;=\; \exp\!\big(QC^\top\big)\; \Delta^\top .
\label{eq:extract}
\end{align}

Applying Lemma~\ref{lem:onehot} with $\phi=\exp$, $U=QC^\top$ and $W=\Delta^\top$ yields Substituting \eqref{eq:extract} into \eqref{eq:att} gives
\begin{equation}
\label{eq:att-cache}
o(Q) \;=\;
\frac{\exp\!\big(QC^\top\big)\; \boxed{\Delta^\top V}}
     {\exp\!\big(QC^\top\big)\; \boxed{\Delta^\top \mathbf{1}_L}} ,
\end{equation}
Here, the \boxed{boxed} terms are independent of $Q$ and can therefore be precomputed and cached offline.  
Observe that $\Delta^\top \in \mathbb{R}^{N\times L}$ and $V \in \mathbb{R}^{L\times d}$, so their product $\Delta^\top V \in \mathbb{R}^{N\times d}$ has no dependence on $L$ in its resulting dimensions.  
This means that the expensive $O(L d)$ aggregation over all events is completed once offline, and only the $O(N d)$ codeword-level aggregates are stored.  
Intuitively, $\Delta$ groups all events that share the same codeword index into $N$ codeword buckets, so online computation only needs to access these bucket-level summaries rather than all $L$ events.

\subsubsection{$E$-tight Output Error Bound}

While \eqref{eq:att-cache} removes online dependence on $L$, we also need to ensure accuracy is controlled.  
Let $E = K - \hat{K}$ with row errors $e_i^\top$. Define
\[
\alpha \;=\; \operatorname{softmax}\!\big(QK^\top\big),
\qquad
\hat{\alpha} \;=\; \operatorname{softmax}\!\big(Q\hat{K}^\top\big),
\]
and let $\delta_i = Q e_i^\top$ be the logit perturbation at position $i$.  
Since softmax is $1$-Lipschitz from $(\ell_\infty,\ell_1)$,
\begin{equation}
\label{eq:softmax-lip}
\|\alpha - \hat{\alpha}\|_1
\;\le\; \|\delta\|_\infty
\;=\; \|Q\|_2 \;\max_i \|e_i\|_2 .
\end{equation}
If each value vector is constrained by a norm bound $|V_i|_2 \le c$ (e.g., via LayerNorm within the model architecture), then
\begin{equation}
\begin{aligned}
\|o(Q) - \hat{o}(Q)\|_2
&= \|(\alpha - \hat{\alpha})^\top V\|_2 \\
&\;\le\; c \;\|\alpha - \hat{\alpha}\|_1 \\
&\;\le\; c \; \|Q\|_2 \;\max_i \|e_i\|_2 .
\end{aligned}
\label{eq:lfree}
\end{equation}

Crucially, the bound in \eqref{eq:lfree} is independent of $L$.  
Moreover, the codebook and commitment losses (Eq.~\ref{eq:vq_loss}) directly act on the reconstruction error $E = K - \hat{K}$ by pulling each codeword toward its assigned keys and simultaneously forcing keys to commit to their selected codewords.  
This bidirectional pressure reduces the worst-case deviation $\max_i \|e_i\|_2$, which appears as the dominant factor in \eqref{eq:lfree}, thereby tightening the output error bound.  
Intuitively, the codebook loss contracts the distance between codewords and their neighborhoods, while the commitment loss prevents unstable switching of assignments, together yielding a more compact quantization space and a provably smaller error constant.

\subsection{Grouped Vector Quantization (GVQ)}
\label{subsec:gvq}

A single small codebook often lacks sufficient capacity to represent diverse feature subspaces.  
One straightforward remedy is to increase the codebook size $N$, but this inevitably enlarges the size of cached results, offsetting the efficiency benefits of quantization.  
To enhance representational capacity without increasing cache size, we replace one large codebook with multiple small codebooks, each dedicated to a disjoint subset of the channel dimension.

However, if we directly apply such grouped quantization to multi-head attention, the number of query heads will be constrained by the number of key-value groups $G$, limiting query expressiveness.  
This observation naturally connects to the idea of \emph{Grouped Query Attention (GQA)}, which decouples the number of query heads $H$ from the number of key-value groups ($H \ge G$).  
Inspired by GQA\cite{gqa}, we propose rouped Vector Quantization (GVQ), which applies vector quantization to multiple small key-value codebooks while preserving independent and potentially more numerous query heads.

We split the channel dimension of $K$ and $V$ into $G$ equal-sized groups, each of dimension $d_g = d/G$:
\begin{align}
K &= [K^1 \| \cdots \| K^G], \quad K^g \in \mathbb{R}^{L \times d_g}, \\
V &= [V^1 \| \cdots \| V^G], \quad V^g \in \mathbb{R}^{L \times d_g}.
\end{align}

Each group $g$ has its own codebook $C^g \in \mathbb{R}^{N \times d_g}$  
and assignment matrix $\Delta^g \in \{0,1\}^{L \times N}$.  

For a query head $Q^h$ mapped to group $g(h)$,  
the attention output is computed as:
\begin{equation}
o^h(Q) = 
\frac{\exp(Q^h {C^{g(h)}}^\top) \; \boxed{\Delta^{g(h)\top} V^{g(h)}}}
     {\exp(Q^h {C^{g(h)}}^\top) \; \boxed{\Delta^{g(h)\top} \mathbf{1}_L}}.
\label{eq:gvq-att}
\end{equation}

Since $d_g = d/G$, the total cache size is:
\[
\sum_{g=1}^G N \, d_g = N \, d,
\]
independent of $H$ and $G$.  
Thus GVQ leverages multiple specialized codebooks to improve representational capacity, while the GQA-inspired head–group decoupling ensures that query flexibility is preserved.

\subsection{Context-Aware Temporal Aggregation}

Ultra-long user histories often span months or even years, and their relevance is strongly modulated by various contextual factors such as temporal gaps, interaction scenarios, or platform characteristics.  

On the query side, contextual features can be integrated in a straightforward manner: as shown in Eq.~(\ref{eq:att-cache}), the cached representations are independent of $Q$, which naturally allows query-side context injection without altering cacheability.  
On the key-value side, behavior-associated signals such as the interaction scenario or absolute timestamps (e.g., the week or hour when an event occurs) can also be seamlessly incorporated into our framework while keeping the caching mechanism intact.  

Among different types of context, relative time emerges as one of the most critical signals, much like positional encodings in large language models. To account for this, we explicitly incorporate relative temporal information into our framework. Inspired by the design of T5\cite{t5}, we adopt a relative position encoding strategy that balances modeling accuracy with cache efficiency.  

We model temporal influence using an exponential kernel:
\begin{equation}
\Phi_\lambda(t_q,t_k)=\exp(-\lambda|t_q-t_k|).
\end{equation}
For $t_q \ge t_k$, this kernel can be factorized as
\begin{equation}
\exp\!\big(-\lambda(t_q - t_k)\big)
= \underbrace{\exp(-\lambda t_q)}_{\text{query-side}}
\cdot
\underbrace{\exp(\lambda t_k)}_{\text{history-side}} = \frac{\Phi_\lambda(t_q,0)}{\Phi_\lambda(0,t_k)}.
\end{equation}
so that all history-side weights are incorporated into offline caches, while only a scalar query-side modulation is applied online.

Let $\Delta \in \{0,1\}^{L\times N}$ be the VQ assignment and $t_k \in \mathbb{R}^L$ the event timestamps.  
For a set of decay rates $\{\lambda_m\}_{m=1}^M$, define
\[
\widetilde{V}_{\lambda_m} = D_{\lambda_m} V, 
\qquad 
\widetilde{\mathbf{1}}_{\lambda_m} = D_{\lambda_m} \mathbf{1},
\]
where $D_{\lambda_m} = \mathrm{diag}\!\big(\exp(\lambda_m t_k)\big)$ are diagonal scaling matrices depending only on historical timestamps.  
The following per-scale, per-codeword aggregates can then be precomputed offline:
\begin{align}
\boxed{\Delta^\top \widetilde{V}_{\lambda_m}} &\in \mathbb{R}^{N \times d}, \\
\boxed{\Delta^\top \widetilde{\mathbf{1}}_{\lambda_m}} &\in \mathbb{R}^{N \times 1}.
\end{align}
These caches depend only on history $(\Delta, t_k, V)$ and remain completely independent of the query.

At inference time, given $t_q$, the query-side factor is $\exp(-\lambda_m t_q)$.  
A small gating module may further produce nonnegative scale weights $\theta_m(Q)$ with $\sum_{m=1}^M \theta_m(Q)=1$.  
Replacing the cache terms in attention then yields
\begin{equation}
o(Q,t_q) \;=\;
\frac{\exp(QC^\top)\; \sum_{m=1}^M \theta_m(Q)\, e^{-\lambda_m t_q}\, \boxed{\Delta^\top \widetilde{V}_{\lambda_m}}}
     {\exp(QC^\top)\; \sum_{m=1}^M \theta_m(Q)\, e^{-\lambda_m t_q}\, \boxed{\Delta^\top \widetilde{\mathbf{1}}_{\lambda_m}}}.
\end{equation}

\subsection{Training \& Inference Strategy}
\label{subsec:inference}

\subsubsection{Training Strategy}
During training, the model jointly optimizes the main recommendation objective and the vector quantization regularization.  
Specifically, the overall loss is defined as
\begin{equation}
\mathcal{L} = \mathcal{L}_{\text{Rec}} + \alpha \sum_{i=1}^G \mathcal{L}^{i}_{\text{VQ}},
\end{equation}
where $\mathcal{L}_{\text{Rec}}$ denotes the recommendation loss (Eq.~\ref{eq:rec_loss}),  
and $\mathcal{L}_{\text{VQ}}$ represents the vector quantization loss (Eq.~\ref{eq:vq_loss}), which includes both the codebook loss and the commitment loss.  
Here, $\alpha$ is a balancing coefficient.

\subsubsection{Inference Strategy}
Both $\boxed{\Delta^\top \cdot \mathbf{1}}$ and $\boxed{\Delta^\top \cdot V}$ can be precomputed offline, and Fig.~\ref{fig:codes} illustrates the difference between our training and inference pipelines. 
In ultra-long sequence modeling, there is an inherent trade-off between efficiency and accuracy. 
To better adapt to practical constraints, we propose three levels of caching strategies that balance storage, update cost, and context compatibility. 

For $\Delta$, we first traverse all items and their side information to build a forward lookup table based on the codebook matrix $C^\top$.
During inference, given a user sequence, the corresponding codes can be retrieved efficiently through table lookup with $O(L)$ complexity. 
This mapping generates a sparse one-hot matrix stored in compressed sparse column (CSC) format with $O(L)$ storage complexity. 
Assuming a system with \( U \) users, \( I \) items, and a codebook size of \( N \), the caching strategies are summarized as follows:

\begin{itemize}
    \item \textbf{VQL$_L$ (light caching).}  
    Only the codes corresponding to each item are cached, leading to a storage complexity of $O(I)$.  
    This strategy is lightweight and update-friendly, but it cannot support context injection since caches contain only static code assignments.

    \item \textbf{VQL$_M$ (medium caching).}  
    The one-hot matrix $\Delta$ is cached in a sparse format, with complexity $O(UL)$.  
    This retains more user-specific information while keeping storage moderate.

    \item \textbf{VQL$_H$ (heavy caching).}  
    Both $\Delta^\top V$ and $\Delta^\top \mathbf{1}$ are cached, with storage complexity $O(UNd)$ and $O(UN)$ respectively.  
    This strategy achieves the highest inference speed and directly supports context-aware extensions, but at the cost of larger memory consumption.
\end{itemize}

In practice, the choice of caching strategy depends on system requirements.  
When frequent model updates are required, VQL$_L$ is preferable since only the codebook centroids need recomputation.  
For platforms with ultra-long user histories and high-dimensional embeddings, VQL$_M$ provides a balanced trade-off between storage overhead and update efficiency.  
When additional memory resources are available, one can adopt VQL$_H$, which maximizes inference efficiency and fully supports context-aware extensions.

\definecolor{codeblue}{RGB}{0,0,180}
\definecolor{codegreen}{RGB}{0,120,0}
\definecolor{codegray}{RGB}{110,110,110}
\definecolor{codeorange}{RGB}{180,90,0}

\lstdefinestyle{topconf}{
    backgroundcolor=\color{white},
    basicstyle=\ttfamily\footnotesize,
    keywordstyle=\color{codeblue},
    commentstyle=\color{codegray}\itshape,
    stringstyle=\color{codeorange},
    numbers=none,
    frame=lines,
    framerule=0.2pt,
    rulecolor=\color{black!20},
    columns=fullflexible,
    keepspaces=true,
    showstringspaces=false,
    aboveskip=4pt,
    belowskip=4pt,
    xleftmargin=4pt,
    xrightmargin=4pt,
    framesep=3pt
}

\begin{figure}[ht!]
\centering
\lstset{style=topconf}
\begin{lstlisting}[language=Python]
# Shapes: B = batch size, L = seq length, N = codebook size, d = dim

# -------- Training stage --------
def VQL_train(Q, C, K, V):
    # Q: [B, d], C: [N, d], K,V: [B, L, d]
    K_vq = vq(K, codebook=C)              # <-- Our key-only VQ
    scores = Q @ K_vq.transpose(1, 2)     # [B, L]
    weights = softmax(scores)             # attention weights
    return weights @ V                    # preserve raw values

# -------- Inference stage --------
def VQL_infer(Q, C, V_cache, Delta_cache):
    # V_cache: [N, d], Delta_cache: [N, 1], precomputed offline
    scores = exp(Q @ C.transpose(1, 2))   # [B, N], L-free
    return (scores @ V_cache) / (scores @ Delta_cache)
\end{lstlisting}
\caption{Unified pseudocode for training and inference in our \baby{} framework. 
Key-only vector quantization enables $L$-independent precomputation of 
V\_cache and Delta\_cache, preserving accuracy while 
removing sequence-length dependence at inference.}
\label{fig:codes}
\end{figure}

\section{Experiments}
\label{sec:experiment}
In this section, we evaluate \baby by comparing it with several long user sequence models to validate our proposed method in terms of both efficiency and inference speed.

\subsection{Experimental Setup} 
\label{sec:setup}

\begin{table}[h]
\caption{Dataset Statistics for Experiments. The table includes the average sequence length (Avg length) and the 90th percentile sequence length (p90 length).}\label{t:dataset}
\begin{tabular}{ccccc}
\toprule
Dataset       & KuaiRand-1K  & KuaiRec & TMALL  \\ 
\midrule
\# Users      & 996  & 7,176      &  13,612  \\
\# Items      & 203,758  & 10,728     & 147,823   \\
\# Interactions & 4,389,984 & 12,530,806    & 1,375,888  \\
\# Features & 36 & 33 & 5  \\
Avg Length      & 4,407 & 1,746     & 124   \\
P90 Length   & 8,879  & 2,912 & 198 \\
\bottomrule
\end{tabular}
\end{table}

\noindent\textbf{Datasets.}
To evaluate the effectiveness of our proposed method, we used two publicly available datasets, which feature extremely long behavior sequences and contain rich side information. The details of the datasets are as follows:

\begin{itemize}
\item \textbf{KuaiRand-1K}\footnote{\url{https://kuairand.com/}}: KwaiRand is a sequential recommendation dataset gathered from the recommendation logs of a video-sharing mobile app. 

\item \textbf{KuaiRec}\footnote{\url{https://kuairec.com/}}: KuaiRec is a real-world dataset collected from the recommendation logs of the Kuaishou video-sharing mobile app.

\item \textbf{Tmall}\footnote{\url{https://tianchi.aliyun.com/dataset/140281}}: 
TMALL is an e-commerce dataset sourced from Alibaba. We use it for CVR
estimation, predicting if a user adds the product to the cart after
a click. 
\end{itemize}

The processed data analysis can be found in Table \ref{t:dataset}. For each dataset, we selected several relevant side information features, which can be categorized into three main types: \textbf{Context Features}, which include static information from the user side, or contextual details such as scene and time within the request; \textbf{Item Features}, which refer to static attributes of the item, including the item ID, category, and brand for products, or tags and duration for short videos; and \textbf{Sequential Information}, which includes the sequence of the static attributes of items, as well as temporal difference information.

\paratitle{Baselines.} To evaluate the effectiveness of \baby, we conduct a comprehensive comparison with 9 state-of-the-art models. 

\begin{itemize}
    \item \textbf{DIN}~\citep{DBLP:conf/kdd/ZhouZSFZMYJLG18} is the most widely used behavior modeling approach, which leverages target attention to model the relationship between candidate items and the short-term behavior sequence.
    \item \textbf{DIEN}~\citep{dien} is an enhanced version of DIN, which introduces a auxialy task for next item prediction.
    \item \textbf{MIMN}~\citep{KDD'19/MIMN}  use a memory-based unit to encode the long user sequence offline.
    \item \textbf{SimSoft}~\citep{CIKM'20/SIM} is a search-based long-term modeling algorithm, where GSU and ESU are trained simultaneously, and the top-k behaviors are chosen via maximum inner product search.
    \item \textbf{ETA}~\citep{arxiv'21/ETA} uses locality-sensitive hashing (LSH) and Hamming distance to accelerate the search process.
    \item \textbf{SDIM}~\citep{CIKM'22/SDIM} gathers behavior items associated with the candidate item with the same hash signature to form the user interest. 
    \item \textbf{TWIN}~\citep{KDD'23/TWIN} unifies the parameters of GSU and ESU, periodically synchronizing the parameters of ESU to GSU.
    \item \textbf{TWIN-V2}~\citep{twin_v2} introduces a hierarchical clustering method that groups items with similar characteristics in life-cycle behaviors for online inference in GSU retrieval.
    \item \textbf{LREA}~\citep{lrea} introduces an end-to-end model that uses linear compression to shorten sequence lengths.
\end{itemize}

\paratitle{Evaluation Metric.} We report AUC and LogLoss as our primary evaluation metrics. 
AUC reflects the ranking quality of the predicted scores, while LogLoss measures the calibration of predicted probabilities. 
It is important to note that in large-scale recommender systems, AUC values are often already very high (e.g., above 0.70–0.80), making further improvements extremely challenging. 
Therefore, even small absolute gains in AUC (e.g., +0.1\%) are regarded as practically significant and can translate to substantial business impact.

\paratitle{Implementations.} To ensure a fair comparison, we adopt the following settings for all methods: the batch size is set to 1024, and the embedding size is set to 64. All models are implemented based on Fuxi-CTR\footnote{https://github.com/reczoo/FuxiCTR}, which provides a robust structure and serves as a baseline for long sequence modeling, as utilized in other studies \cite{xu2024multi}. The models are trained using an early stopping mechanism, with the learning rate adjusted within the range of [1e-3, 3e-4, 1e-4]. To maintain experimental fairness, we optimize all baselines according to their original papers.

For the KuaiRand-1K dataset, all long sequence modeling methods are applied to the most recent 5000 items, which are subsequently compressed into sequences of length 100. For short sequence modeling, we use the most recent 100 items. In the KuaiRec dataset, long sequence modeling considers the most recent 2000 items, compressed to sequences of length 100, while short sequence modeling also uses 100 items. For the TMALL dataset, we select sequences with a maximum length of 300, compressing them to sequences of length 50, and use the most recent 50 items for short sequence modeling. To ensure a fair comparison, we adjust only the long sequence modeling parts. For long sequence modeling methods that do not support relative time, absolute position encoding is introduced. All prediction networks employ a three-layer DNN structure, with hidden layers of sizes [32, 16].

\begin{table*}[h]
\centering
\caption{The performance comparison between the baselines and \baby is presented. For each column, the best result is highlighted in bold, and the second-best is \underline{underlined}. Statistical significance is evaluated at the 0.05 level.}
\label{tab:overall}
\begin{tabular}{ccccccc}
\toprule
 & \multicolumn{2}{c}{\textbf{KuaiRand-1K}} & \multicolumn{2}{c}{\textbf{KuaiRec}} & \multicolumn{2}{c}{\textbf{TMALL}} \\
\cmidrule(lr){2-3} \cmidrule(lr){4-5} \cmidrule(lr){6-7}
 & \textbf{AUC\;($\uparrow$)} & \textbf{LogLoss\;($\downarrow$)}
 & \textbf{AUC\;($\uparrow$)} & \textbf{LogLoss\;($\downarrow$)}
 & \textbf{AUC\;($\uparrow$)} & \textbf{LogLoss\;($\downarrow$)} \\
\midrule
\textbf{DIN}     & 67.64 & 65.08 & 79.78 & 50.99 & 71.26 & 32.57 \\
\textbf{DIEN}    & 67.69 & 65.00 & 79.57 & 51.01 & 71.29 & 32.54 \\
\textbf{MIMN}    & 67.81 & 64.75 & 79.88 & 50.97 & 71.20 & 32.59 \\
\textbf{SimSoft} & 68.64 & 65.04 & \underline{80.33} & 50.83 & 71.30 & 32.56 \\
\textbf{ETA}     & 67.91 & 64.64 & 80.32 & 50.78 & 71.33 & 32.55 \\
\textbf{SDIM}    & 68.88 & 63.97 & 79.94 & 51.21 & 71.18 & 32.61 \\
\textbf{TWIN}    & 68.98 & 64.57 & 80.15 & \underline{50.64} & 71.33 & 32.54 \\
\textbf{TWIN-V2} & 70.02 & 63.84 & 79.99 & 51.47 & \underline{71.35} & \underline{32.53} \\
\textbf{LREA}    & \underline{70.64} & \underline{62.97} & 80.21 & 50.65 & 71.22 & 32.58 \\
\textbf{\baby}   & \textbf{71.47} & \textbf{62.44} & \textbf{80.59} & \textbf{50.42} & \textbf{71.50} & \textbf{32.49} \\
\bottomrule
\end{tabular}
\end{table*}

\subsection{Overall Performance} 
\label{sec:performance}
As shown in Table~\ref{tab:overall}, Compared to models like DIN and DIEN, which can only model very short behavior sequences, they achieved almost the worst performance. MIMN introduced a sequence encoder, allowing it to model complete behavior sequences and achieve some improvements. However, its performance is limited by the late-stage interaction between the sequence and features, restricting further enhancement. For the two-stage models ESU and GSU, SimSoft is the strongest baseline on the KuaiRec dataset, indicating that top-k-based methods still have potential. However, on the longer KwaiRand dataset, LREA performs better due to its ability to directly compress behavior sequences and retain more information, overcoming the limitations of top-k. Our proposed \baby model achieved the best results across all datasets. This is because we only quantized K and directly supervised the loss using VQ's loss. Additionally, we designed multiple context injection methods. These strategies effectively balance low compression loss, context awareness, and low inference latency.

\subsection{Ablation Study}
In this section, we conduct experiments to assess the impact of different variants of \baby on the results.

\subsubsection{Effectiveness of the Vector Quantization Strategy}
In \baby, we integrate end-to-end vector quantization techniques for modeling ultra-long sequences, supported by both codebook loss and commitment loss. To verify the effectiveness of these components, we design three ablation variants for comparison:

\begin{itemize}
    \item $\rm{\baby}_{\neg{NS}}$: We remove the VQ loss, meaning that the assignment of keys to their nearest codewords is no longer explicitly supervised.  
    \item $\rm{\baby}_{TS}$: We first train a full long-attention model and then employ a VQ-VAE model to cluster items offline for inference.  
    \item $\rm{\baby}_{KV}$: Unlike \baby, which quantizes only the keys, this variant quantizes both keys and values, thereby introducing additional distortion.  
\end{itemize}

\begin{table}[htbp]
\small
\centering
\caption{Performance comparison of vector quantization variants on two datasets (all metrics in percentage).}
\label{tb:ab_1}
\begin{tabular}{lccccc}
\toprule
\textbf{Dataset} & \textbf{Metric} & $\rm{\baby}_{\neg{NS}}$ & $\rm{\baby}_{TS}$ & $\rm{\baby}_{KV}$ & \textbf{VQL} \\ 
\midrule
\multirow{2}{*}{KuaiRand-1K} 
  & AUC & 70.52 & 70.83 & 69.14 & \textbf{71.47} \\ 
  & LogLoss & 63.15 & 62.81 & 63.99 & \textbf{62.44} \\ 
\midrule
\multirow{2}{*}{TMALL} 
  & AUC & 71.33 & 71.40 & 71.20 & \textbf{71.50} \\ 
  & LogLoss & 32.54 & 32.62 & 32.59 & \textbf{32.49} \\ 
\bottomrule
\end{tabular}
\end{table}

As shown in Table~\ref{tb:ab_1}, removing the VQ loss makes the assignment of keys to codewords non-differentiable, so the quantization process is no longer explicitly supervised, leading to degraded performance. The two-stage learning approach, compared to the one-stage end-to-end model, introduces discrepancies between training and inference, which also harms accuracy. Finally, when both $K$ and $V$ are quantized, performance drops substantially: quantizing $K$ only influences the attention weights over items, whereas quantizing $V$ directly distorts the item representations being aggregated, causing the model to lose fine-grained semantic information.

\subsubsection{Effectiveness of the GVQ and Context-Aware Temporal Aggregation}
In \baby, we design two mechanisms for injecting context information, and here we evaluate their effectiveness through three ablated variants:
\begin{itemize}
    \item $\rm{\baby}_{\neg{C}}$: Dynamic context information (e.g., absolute time, scene features) is removed, making the model compatible with the VQL$_L$ caching strategy.  
    \item $\rm{\baby}_{\neg{G}}$: The grouped vector quantization (GVQ) mechanism is removed, so all query context is modeled as a single query through single-head attention.  
    \item $\rm{\baby}_{\neg{T}}$: The temporal aggregation module is removed, so relative time information is no longer captured.  
\end{itemize}

\begin{table}[htbp]
\centering
\caption{Performance evaluation of context injection strategies across two datasets (all metrics in percentage).}
\label{tb:context}
\begin{tabular}{lccccc}
\toprule
\textbf{Dataset} & \textbf{Metric} & $\rm{\baby}_{\neg{C}}$ & $\rm{\baby}_{\neg{G}}$ & $\rm{\baby}_{\neg{T}}$ & \textbf{VQL} \\ 
\midrule
\multirow{2}{*}{KuaiRand-1K} 
  & AUC & 71.35 & 71.16 & 71.28 & \textbf{71.47} \\ 
  & LogLoss & 62.29 & 62.18 & 62.84 & \textbf{62.44} \\ 
\midrule
\multirow{2}{*}{KuaiRec} 
  & AUC & 80.42 & 80.41 & 80.37 & \textbf{80.52} \\ 
  & LogLoss & 50.67 & 50.69 & 50.75 & \textbf{50.42} \\ 
\bottomrule
\end{tabular}
\end{table}

As shown in Table~\ref{tb:context}, removing context information from the sequence degrades performance, since rich contextual signals help capture comprehensive user interests. Eliminating GVQ also reduces accuracy, as multi-codebook grouping plays a role similar to multi-head attention by enabling more nuanced modeling of interest subspaces. Finally, removing the temporal aggregation mechanism causes further decline, because relative time-awareness is crucial for fine-grained modeling of temporal gaps, whereas absolute time or order alone can only capture coarse-grained or periodic patterns.





\subsection{Scaling with Long Sequence Length}
\label{sec:scaling}
\begin{figure}
    \centering
    \includegraphics[scale=0.14]{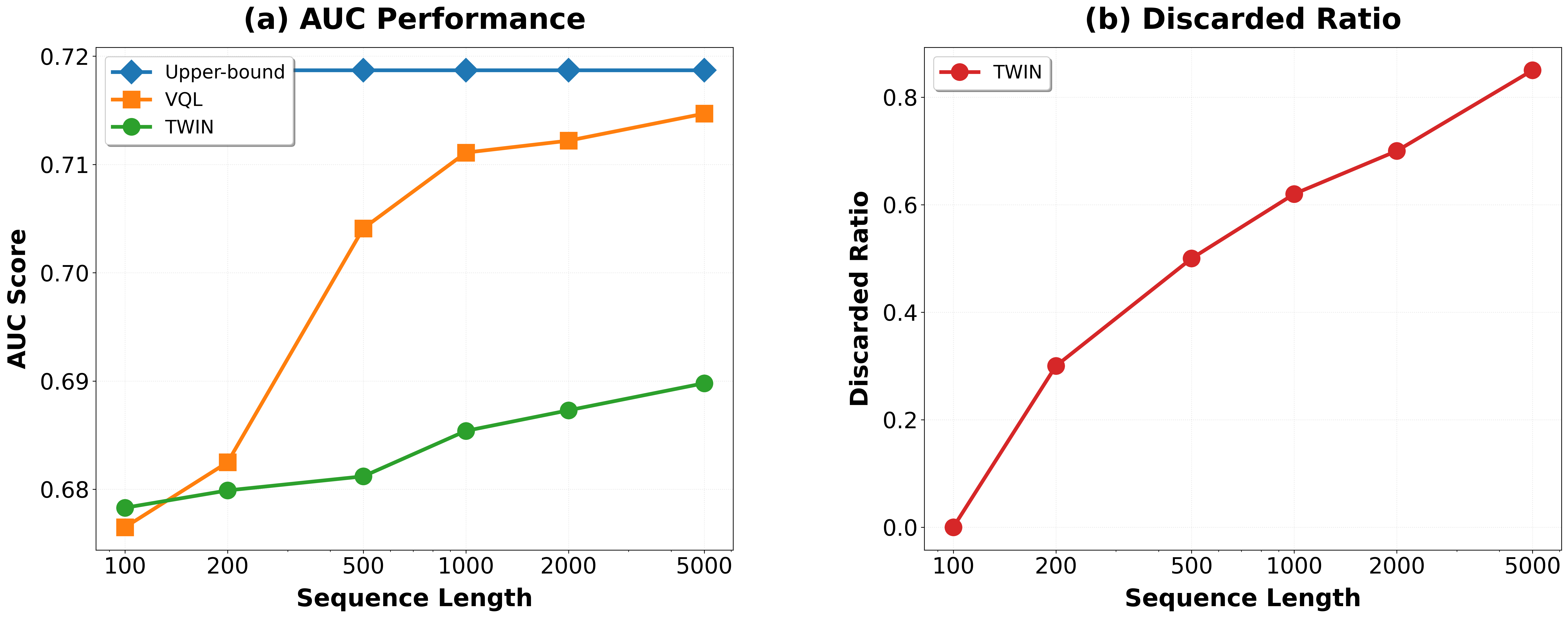}
    \caption{
    Performance comparison under different sequence lengths \(L\).  
    }
    \label{fig:seq_len}
\end{figure}

We further investigate how extending the long-sequence length \(L\) impacts model performance while fixing the short-sequence modeling window to 100 steps.  
Experiments are conducted on the KuaiRand-1K dataset, with \(L\) varied in \([100, 200, 500, 1000, 2000, 5000]\).  
We focus on comparing our method with top-\(k\) retrieval approaches such as TWIN.  
To provide a reference point, we additionally introduce a full-length attention baseline, which serves as an upper-bound performance ceiling.  
The results are illustrated in Figure~\ref{fig:seq_len}.  

Interestingly, when \(L=100\), TWIN slightly outperforms VQL.  
This is expected since the top-\(k\) mechanism in TWIN is effectively deactivated (reducing to standard self-attention), whereas VQL still incurs a small quantization loss from compressing item embeddings.  
However, as \(L\) grows, VQL exhibits consistent and rapid performance improvements, surpassing TWIN once \(L \geq 200\).  
This trend highlights the key limitation of top-\(k\) retrieval: the accuracy degradation grows with longer sequences.  

To better quantify this effect, we examine the proportion of discarded attention weights—i.e., the ratio of pruned items \((L-k)\) to the total sequence length \(L\)—as shown in the right panel of Figure~\ref{fig:seq_len}.  
The monotonic growth of this ratio confirms our hypothesis from the introduction: top-\(k\) compression inevitably causes substantial information loss when \(L \gg k\).  
In contrast, VQL maintains stable performance by keeping compression loss length-independent, making it more suitable for ultra-long sequence modeling. 

\subsection{Analysis of the Computational Efficiency}
\begin{figure}
    \centering
    \includegraphics[scale=0.14]{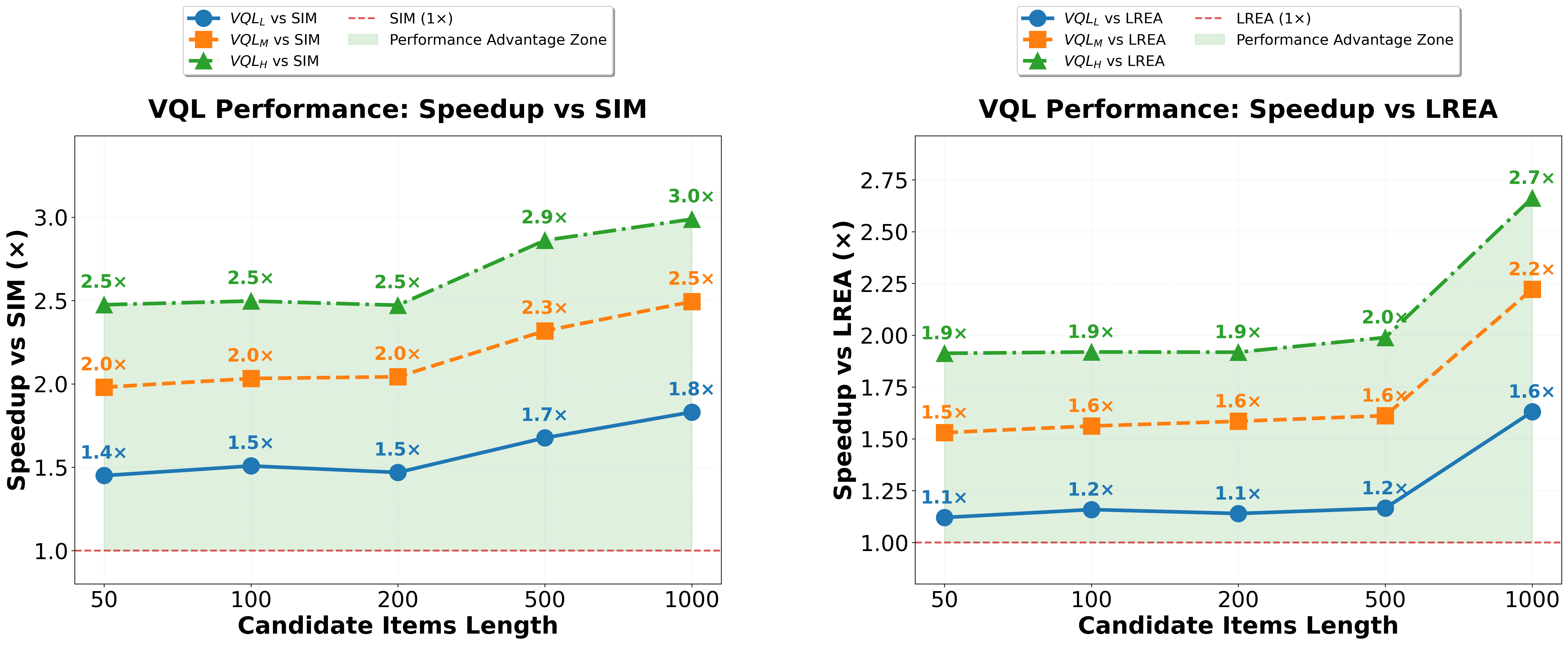}
    \caption{Average response time of different inference strategies in KuaiRec Dataset.}
    \label{fig:time_comp}
\end{figure}

For long-period sequence modeling, reducing inference costs is crucial due to the constraints of online inference resources. In Section \ref{subsec:inference}, we discussed some caching strategies and we also aim to evaluate the computational efficiency of our model. In a real online system, a single inference responds to one user's request, meaning that a batch will contain \(m\) candidate items, with identical user-side information. Due to the lack of sufficient candidate items in offline datasets, we select the last \(m\) items from a user's complete long-period sequence as candidate items and truncate the sequence to the most recent 2000 items. The number of candidate items is set to [50, 100, 200, 500, 1000]. Our inference performance is tested on a machine configured with Intel(R) Xeon(R) Platinum 8352Y CPU @ 2.20GHz ×2 and NVIDIA A800-SXM4-80GB × 8 GPUs. We run the inference 10 times and take the average. The average response time of the final model is shown in Figure \ref{fig:time_comp}.

As shown in the Figure \ref{fig:time_comp}. All three caching strategies of VQL achieved the fastest inference speed. As the cached content gradually increases, the inference delay significantly decreases. Since VQL is independent of the compressed sequence and query, there is no significant change in inference delay even with a substantial increase in candidate items. SIM has the lowest computational efficiency because it directly retrieves top-k items, which is very time-consuming. Its delay significantly increases when the candidate items reach 200. LREA uses target-attention from DIN and requires additional MLP processing. Although it is not affected by candidate items at short lengths, its performance significantly declines when the candidates reach 1000.

\section{Conclusion}

In this work, we present \textbf{VQL}, an end-to-end vector quantization attention framework for ultra-long user behavior modeling.  
VQL addresses the challenges of balancing efficiency, accuracy, and context-awareness through four core advantages.  
First, \emph{K-only quantization} compresses only the keys while preserving raw values, avoiding value-side distortion and retaining the fidelity of the original attention structure.  
Second, the \emph{L-Free caching} property, enabled by the one-hot extraction identity, precomputes all key–value aggregates offline into $L$-independent caches, eliminating online dependence on sequence length.  
Third, \emph{E-Tight quantization} establishes an error bound directly proportional to the quantization error and independent of $L$, with multi-codebook grouping further tightening this bound.  
Finally, \emph{Q-Free context injection} supports diverse context signals—such as static attributes and temporal dynamics—without enlarging the codebook or breaking cache compatibility.  

Extensive experiments on three large-scale datasets demonstrate that VQL consistently outperforms strong baselines.
Beyond academic benchmarks, we have also validated VQL in our industrial offline environment, where it yielded consistent AUC gains.  
Online A/B testing is currently underway, and in future work we will report the realized online revenue lift and describe in detail the engineering practices that enable full deployment in production recommender systems.

\newpage

\bibliographystyle{ACM-Reference-Format}
\bibliography{sample-base}

\end{document}